\documentclass[prl,twocolumn,a4paper,groupedaddress,showpacs,floatfix,aps,10pt,
longbibliography,nobibnotes]{revtex4-1}
\usepackage{graphicx,amsmath,amssymb,amsfonts,dsfont,subfigure,color}

\newcommand{\ket}[1]{\ensuremath{|#1\rangle}}
\newcommand{\mc}[1]{\ensuremath{\mathcal{#1}}}

\usepackage{color}

\newcommand{\vroW}{\varrho}

\begin{document}

\title{Coherent Microwave-to-Optical Conversion via Six-Wave Mixing in Rydberg Atoms}

\author{Jingshan Han${}^{1}$}
\author{Thibault  Vogt${}^{1,2}$}
\author{Christian Gross${}^{1}$}
\author{Dieter Jaksch${}^{3,1}$}
\author{Martin Kiffner${}^{1,3}$}
\author{Wenhui Li${}^{1,4}$}

\affiliation{Centre for Quantum Technologies, National University of Singapore, 3 Science Drive 2, Singapore 117543${}^1$}
\affiliation{MajuLab, CNRS-UNS-NUS-NTU International Joint Research Unit UMI 3654, Singapore 117543${}^2$}
\affiliation{Clarendon Laboratory, University of Oxford, Parks Road, Oxford OX1 3PU, United Kingdom${}^3$}
\affiliation{Department of Physics, National University of Singapore, 117542, Singapore${}^4$}

\pacs{42.50.Gy,42.65.Ky,32.80.Ee}

%
%
%

\begin{abstract}
We present an experimental demonstration of converting a microwave field to an
optical field via frequency mixing in a cloud of cold $^{87}$Rb atoms, where the
microwave field strongly couples to an electric dipole transition between
Rydberg states. We show that the conversion allows the phase information of the
microwave field to be coherently transferred to the optical field. With the
current energy level scheme and experimental geometry, we achieve a photon
conversion efficiency of $\sim$ 0.3\% at low microwave intensities and a broad
conversion bandwidth of more than 4~MHz. Theoretical simulations agree well with
the experimental data, and indicate that near-unit efficiency is possible in
future experiments.
\end{abstract}

\maketitle

Coherent and efficient conversion from microwave and terahertz radiation into
optical fields and vice versa has tremendous potential for developing
next-generation classical and quantum technologies. For example, these methods
would facilitate the detection and imaging of millimeter waves with various
applications in medicine, security screening and
avionics~\cite{adam:11,chan:07,tonouchi:07,Zhang2017}.
In the quantum domain, coherent microwave-optical conversion is essential for
realizing quantum hybrid systems~\cite{Xiang2013} where spin systems or
superconducting qubits are coupled to optical photons that can be transported
with low noise in optical fibers~\cite{Kimble2008}.
The challenge in microwave-optical conversion is to devise a suitable platform
that couples strongly to both frequency bands, which are separated by several
orders of magnitude in frequency, and provides an efficient link between them.
Experimental work on microwave-optical conversion has been based on
ferromagnetic magnons~\cite{Hisatomi2016}, frequency mixing in $\Lambda$-type
atomic ensembles~\cite{Williamson:2014,OBrien:2014,Blum:2015,Hafezi:2012},
whispering gallery resonators~\cite{Strekalov2009a,Rueda2016}, or nanomechanical
oscillators~\cite{Bochmann2013,Andrews:2014,Bagci:2014}.  All of these schemes
include cavities to enhance the coupling to microwaves. The realization of near-unit
conversion efficiencies as e.g. required for transmitting quantum information
remains an outstanding and important goal.
Recently, highly excited Rydberg atoms have been identified as
a promising alternative~\cite{kiffner:16,jacobs:17} as they feature strong electric dipole
transitions in a wide frequency range from microwaves to terahertz~\cite{gallagher:ryd}.

In this letter, we demonstrate coherent microwave-to-optical conversion of
classical fields via six-wave mixing in Rydberg atoms. Due to the strong
coupling of millimeter waves to Rydberg transitions, the conversion is realized
in free space. In contrast to
millimeter-wave induced optical fluorescence~\cite{Wade2016}, frequency mixing
is employed here to convert a microwave field into a unidirectional single
frequency optical field. The long lifetime of Rydberg states allows us to make
use of electromagnetically induced transparency
(EIT)~\cite{mohapatra2007coherent}, which significantly enhances the conversion
efficiency~\cite{fleischhauer:05}. A free-space photon-conversion efficiency of
0.3\% with a bandwidth of more than 4~MHz is achieved with our current experimental
geometry. Optimized geometry and energy level configurations should enable the
broadband inter-conversion of microwave and optical fields with near-unit
efficiency~\cite{kiffner:16}. Our results thus constitute a major step
towards using Rydberg atoms for transferring quantum states between optical and
microwave photons.
%
\begin{figure}[t!]
\begin{center}
\includegraphics[width=8cm]{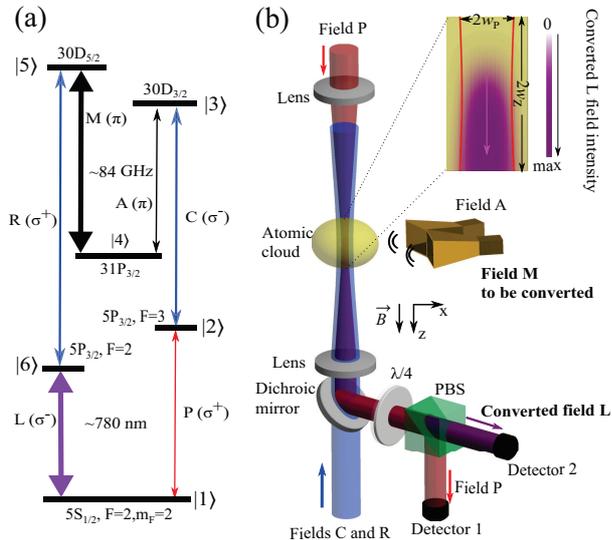}
\end{center}
\caption{\label{fig1}
(a) Relevant energy levels of a ${}^{87}$Rb atom coupled
by six nearly-resonant electromagnetic fields: $\ket{1} =
|5S_{1/2}, F=2, m_F=2\rangle$, $\ket{2} = |5P_{3/2}, F=3, m_F=3\rangle$,
$\ket{3} = |30D_{3/2}, m_J= 1/2\rangle$, $\ket{4} = |31P_{3/2}, m_J=
1/2\rangle$, $\ket{5} = |30D_{5/2}, m_J= 1/2\rangle$, and $\ket{6} = |5P_{3/2},
F=2, m_F= 1\rangle$. Polarizations of the fields are indicated in brackets.
The microwave field M ($\approx 84\ $GHz) is converted to the light field L ($\approx 780 \ $nm)
by six-wave mixing.
(b) Experimental setup. Auxiliary light fields P, C, and R propagate
collinearly along the $z$ axis. They are focused onto the center
of a Gaussian-distributed atomic cloud. The fields M and A
are emitted from horn antennas enclosing an angle of 20$^o$, and
propagate horizontally. The bias magnetic field B along $z$ defines the
quantization axis. The co-propagating fields L and P are separated by a polarization splitter ($\lambda/4$ + PBS) and
detected simultaneously with avalanche photodiodes. The inset shows the simulated intensity and beam profile of the L field.
}
\end{figure}
%

The energy levels for the six-wave mixing are shown in Fig.~\ref{fig1}(a), and
the experimental setup is illustrated in Fig.~\ref{fig1}(b). The conversion of
the input microwave field M into the optical field L is achieved via frequency
mixing with four input auxiliary fields P, C, A, and R in a cold atomic cloud.
Starting from the spin polarized ground state $|1\rangle$, the auxiliary fields
and the microwave field M, all of which are nearly resonant with the
corresponding atomic transitions, create a coherence between
the states $|1\rangle$ and $|6\rangle$. This induces the emission of the light
field L with frequency $\omega_{\text{L}}= \omega_{\text{P}} + \omega_{\text{C}}
- \omega_{\text{A}} + \omega_{\text{M}} - \omega_{\text{R}}$ such that the
resonant six-wave mixing loop is completed, where $\omega_{\text{X}}$ is the
frequency of field X
($\text{X}\in\{\text{P},\text{R},\text{M},\text{C},\text{L},\text{A}\}$). The
emission direction of field L is determined by the phase matching
condition $\mathbf{k}_{\text{L}} = \mathbf{k}_{\text{P}} + \mathbf{k}_{\text{C}}
- \mathbf{k}_{\text{A}} + \mathbf{k}_{\text{M}} - \mathbf{k}_{\text{R}}$, where
$\mathbf{k}_\text{X}$ is the wave vector of the corresponding field. The wave
vectors of the microwave fields $\mathbf{k}_{\text{A}}$ and  $\mathbf{k}_{\text{M}}$
are negligible since they are much smaller than
those of the optical fields and to an excellent approximation, they cancel each
other. Moreover, we have $\mathbf{k}_{\text{C}} \approx \mathbf{k}_{\text{R}}$,
thus the converted light field L propagates in the same direction as the input
field P. The transverse profile of the converted light field L resembles that of
the auxiliary field P due to pulse matching~\cite{harris:93,harris:94} as
illustrated in Fig.~\ref{fig1}(b).

An experimental measurement begins with the preparation of a cold cloud of
$^{87}$Rb atoms in the $|5S_{1/2}, F=2, m_F=2\rangle$ state in a magnetic field
of 6.1 G, as described previously in~\cite{han2015lensing}. At this stage,
the atomic cloud has a temperature of about $70\ \mu\text{K}$, a $1/e^2$ radius
of $w_z=1.85(10) \, \textrm{mm}$ along the $z$ direction, and a peak atomic
density $n_0 = 2.1 (2) \times 10^{10} \, \textrm{cm}^{-3}$. We then switch on
all the input laser and microwave fields simultaneously for frequency mixing.
The beams for both C and R fields are derived from a single 482 nm laser, while
that of the P field comes from a 780 nm laser, and the two lasers are frequency
locked to a single high-finesse temperature stabilized Fabry-Perot cavity
\cite{han2015lensing}. The $1/e^2$ beam radii of these Gaussian fields at the
center of the atomic cloud are $w_{\text{P}}=25(1)\ \mu \text{m}$,
$w_{\text{C}}=54(2)\ \mu \text{m}$, and $w_{\text{R}}=45(1)\ \mu \text{m}$,
respectively; and
their corresponding peak Rabi frequencies are
$\Omega^{(0)}_{\text{P}}=2 \pi \times1.14(7)\ \text{MHz}$, $\Omega^{(0)}_{\text{C}}=2 \pi \times9.0(5)
\ \text{MHz}$, and $\Omega^{(0)}_{\text{R}}=2 \pi \times6.2(3) \ \text{MHz}$. The two
microwave fields M and A, with a frequency separation of around 450 MHz, are
generated by two different microwave sources via frequency multiplication. They
are emitted from two separate horn antennas, and propagate in the horizontal
plane through the center of the atomic cloud, as shown in Fig.~\ref{fig1}(b).
The Rabi frequencies $\Omega_{\text{M}}$ and $\Omega_{\text{A}}$ are
approximately uniform across the atomic cloud volume that intersects the laser
beams. The Rabi frequency of the A field is $\Omega_{\text{A}}=2 \pi \times1.0(1) \
\text{MHz}$, while the Rabi frequency of the M field $\Omega_{\text{M}}$ is
varied in different measurements. The details of the microwave Rabi frequency
calibrations are presented in~\cite{Note1}. The P and L fields that emerge
from the atomic cloud are collected by a
diffraction-limited optical system~\cite{han2015lensing}, and separated using a
quarter-wave plate and a polarization beam splitter (PBS). Their respective powers
are measured with two different avalanche photodiode detectors. Each optical power measurement is an average of the recorded time-dependent signal in the
range from 6 to $16\ \mu$s  after switching on all the fields simultaneously,
where the delay ensures the steady state is fully reached.

%
\begin{figure}[t!]
\begin{center}
\includegraphics[width=8.5cm]{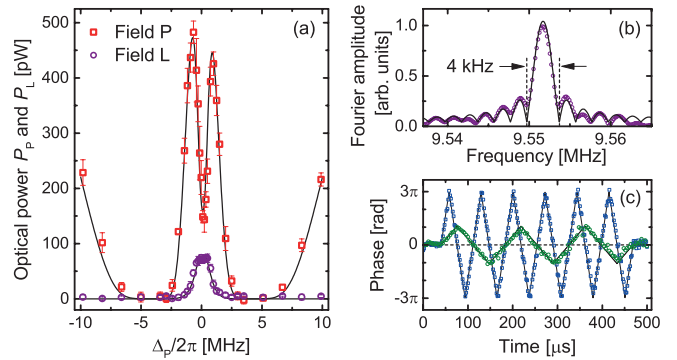}

\end{center}
\caption{\label{fig2}
(a) Spectra of fields P (red squares) and L (purple circles).
The microwave field M has detuning $\Delta_{\text{M}}= 0$ and Rabi
frequency $\Omega_{\text{M}}=2 \pi \times 1.25(12)\ \text{MHz}$. The error bars indicate the
standard deviation of 5 measurements. The solid curves
are obtained from our theoretical model.
(b) Fourier-transform of the optical heterodyne signal between the L field and a
reference optical field for a pulse duration of $500 \, \mu\mathrm{s}$. The
frequency difference between the two fields is $f_c = 9.5517 \, \mathrm{MHz}$. The
solid line shows the fit of a $|$sinc$|$ function to the data.
(c) Relative phase of the heterodyne signals for a phase modulated M field. Triangular
modulations of $ 7 \, \mathrm{kHz}$ and $\pi$ amplitude (green circles) and of $ 14 \, \mathrm{kHz} $
and 3$\pi$ amplitude (blue squares) are shown. The phase is extracted by numerically demodulating the
heterodyne signals~\cite{Note1}. The solid lines show the input phase modulation of the M field.
}
\end{figure}
%
%
We experimentally demonstrate the coherent microwave-to-optical conversion via
the six-wave mixing process by two measurements. First, we scan the detuning $\Delta_{\text{P}}$ of
the P field  across the atomic resonance and measure the power of the transmitted field P ($P_{\text{P}}$),
and the power of the converted optical field L ($P_{\text{L}}$) simultaneously.
All other input fields are held on resonance.
The results of this measurement are shown in Fig.~\ref{fig2}(a),
where the spectrum of the transmitted  field P  (red squares) exhibits a double peak structure.
The signature of the six-wave mixing process is the converted field L (purple circles), and its spectrum
features a pronounced peak around $\Delta_{\text{P}}=0$.

Second, to verify the coherence of the conversion, we perform
optical heterodyne measurements between the L field and a reference field that
is derived from the same laser as the P field. Fig.~\ref{fig2}(b) shows that the
Fourier spectrum of  a 500$\ \mu$s long beat note signal has a transform limited
sinc function dependence. The central frequency of the spectrum confirms that
the frequency of the converted field L is determined by the resonance
condition for the six-wave mixing process. Furthermore, we phase modulate the M
field with a triangular modulation function and observe the recovery of the
phase modulation in the optical heterodyne measurements, as shown in
Fig.~\ref{fig2}(c). This demonstrates that the phase information is coherently
transferred in the conversion, as expected for a nonlinear frequency mixing
process.

%
\begin{figure}[t!]
\begin{center}
\includegraphics[width=8.5cm]{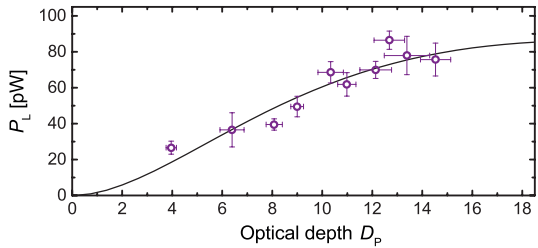}
\end{center}
\caption{\label{fig3}
$P_{\text{L}}$ vs. optical depth $D_{\text{P}}$, with all
fields on resonance. $D_{\text{P}}$ is varied by changing $n_0$, and the other conditions are
the same as for Fig.~\ref{fig2}(a). The circles are experimental data, and the solid line is
the theoretically simulated curve. The error bars correspond to
the standard deviation of 4 measurements.
}
\end{figure}
%
%
We simulate the experimental spectra by modelling the interaction of the laser
and microwave fields with the atomic ensemble within the framework of coupled
Maxwell-Bloch equations~\cite{Note1}. The time evolution of the atomic density
operator $\vroW$ is given by a Markovian master equation ($\hbar$ is  the reduced Planck constant),
\begin{equation}
\partial_t \vroW = - \frac{i}{\hbar} [ H , \vroW ]
+\mc{L}_{\gamma}\vroW+\mc{L}_{\text{deph}}\vroW\,,
\label{master_eq}
\end{equation}
where $H$ is the Hamiltonian describing the interaction of an independent
atom with the six fields, and the term $\mc{L}_{\gamma}\vroW$ describes
spontaneous decay of the excited states. The last term
$\mc{L}_{\text{deph}}\vroW$ in Eq.~(\ref{master_eq}) accounts for dephasing of
atomic coherences involving the Rydberg states $\ket{3}$, $\ket{4}$, and
$\ket{5}$ with the dephasing rates $\gamma_{d}$,  $\gamma_{DD}$, and
$\gamma_{d'}$, respectively~\cite{Note1}. The sources of decoherence are the finite laser linewidths, atomic collisions, and dipole-dipole interactions
between Rydberg atoms.
The dephasing rates affect the P and L spectra and are found
by fitting the steady state solution of coupled Maxwell-Bloch equations to the experimental spectra in Fig.~\ref{fig2}(a).
All other parameters are taken from independent
experimental measurements and calibrations. We obtain $\gamma_{d} = 2 \pi \times $150 kHz,  $\gamma_{DD} = 2 \pi
\times$ 150 kHz and $\gamma_{d'} = 2 \pi \times$ 560 kHz and keep these values
fixed in all simulations.

The system in Eq.~(\ref{master_eq}) exhibits an approximate dark state~\cite{Note1}
\begin{align}
\ket{D}  \propto
\left(\Omega_{\text{M}}^*\Omega_{\text{C}}^*\ket{1}-\Omega_{\text{M}}^*\Omega_{\text{P}}\ket{3}
+\Omega_{\text{A}}^*\Omega_{\text{P}}\ket{5}\right)\label{dark}
\end{align}
for
$\Omega_{\text{L}}/\Omega_{\text{P}}=-\Omega_{\text{A}}^*\Omega_{\text{R}}^*/(\Omega_{\text{M}}^*\Omega_{\text{C}}^*)$,
where $\Omega_{\text{L}}$ is the Rabi frequency of field L.
This state has non-zero population only in
metastable states $\ket{1}$, $\ket{3}$, and $\ket{5}$, and is decoupled from all
the fields. The population in $\ket{D}$ increases with the build-up of the converted light
field along the $z$ direction, and thus $P_{\text{L}}$ saturates when all atoms are trapped in this state. Fig.~\ref{fig3} shows the dependence of the output
power $P_{\text{L}}$ on the optical depth $D_{\text{P}}\propto n_0 w_z$ of the atomic cloud, and the theory curve agrees well with the experimental data.
The predicted saturation at $D_{\text{P}}\approx 20$ is consistent with the population in $\ket{D}$ exceeding 99.8\% at this optical depth.
%

\begin{figure}[t!]
\begin{center}
\includegraphics[width=8.5cm]{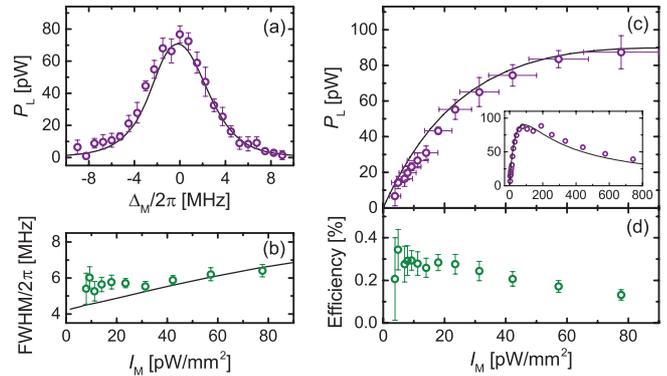}
\end{center}
\caption{\label{fig4}
(a) Spectrum of $P_{\text{L}}$ against microwave detuning
$\Delta_{\text{M}}$ for $I_{\text{M}} = 42 (8)	\ \text{pW/mm}^2$.
(b) FWHM vs. $I_{\text{M}}$. The FWHM is extracted by fitting
a squared Lorentzian function to the spectra.
(c) $P_{\text{L}}$ as a function of M field intensity
$I_{\text{M}}$. The inset shows $P_{\text{L}}$ over a much larger range of $I_{\text{M}}$.
(d) Efficiency $\eta$ calculated for the data shown in (c).
Circles represent experimental data and solid curves simulation results. Vertical error bars in (a) and (c) correspond to the standard deviation of 4-6
measurements, and in (b) to the errors from fitting. Horizontal error bars in (c) are estimated uncertainties. The error bars in (d) are calculated from those in (c).
}
\end{figure}
%

%
Next we analyze the dependence of the conversion process on detuning and
intensity of the microwave field M. All auxiliary fields are kept on
resonance and at constant intensity. Fig.~\ref{fig4}(a) shows $P_{\text{L}}$ as a function of the microwave detuning
$\Delta_{\text{M}}$. We find that the spectrum of the L field
can be approximated by a squared Lorentzian function centered at $\Delta_{\text{M}}=0$,
and its full width at half maximum (FWHM) is $\approx 6\ \text{MHz}$.
The FWHM extracted from microwave spectra at
different intensities $I_\text{M}$ is plotted in Fig.~\ref{fig4}(b). The FWHM has a finite value $> 4$~MHz in the low intensity limit, and increases slowly
with $I_\text{M}$ due to power broadening.
This large bandwidth is one of the distinguishing features of our scheme
and is essential for extending the conversion scheme to the single photon
level~\cite{Rueda2016}.
In Fig.~\ref{fig4}(c), we show measurements of $P_{\text{L}}$ vs.~the intensity of the microwave field $I_\text{M}$ at
$\Delta_{\text{M}} = 0$. We find that the converted power $P_{\text{L}}$ increases  approximately linearly at
low microwave intensities, and thus our conversion scheme is expected to work in the limit of very weak input fields.
The  decrease of $P_{\text{L}}$  at large intensities arises
because the six-wave mixing process becomes inefficient if the Rabi frequency $\Omega_{\text{M}}$ is
much larger than the Rabi frequency  $\Omega_{\text{A}}$ of the auxiliary microwave.
All the theoretical curves in Fig.~\ref{fig4} agree well with the experimental data.

We evaluate the photon conversion efficiency of our
setup by considering the cylindrical volume ${\mathcal{V}}$ where the atomic cloud
and all six fields overlap. This volume has a diameter $\sim 2 w_{\text{P}}$
and a length $\sim 2 w_z$ [see Fig.~\ref{fig1}(b)].
We define the conversion efficiency as
\begin{align}
\eta = \frac{P_{\text{L}}/\hbar \omega_{\text{L}}}{  I_{\text{M}} S_{\text{M}}/ \hbar \omega_{\text{M}}},
\label{eta}
\end{align}
where $S_{\text{M}} = 4 w_{\text{P}} w_z$ is the cross-section of the volume ${\mathcal{V}}$ perpendicular to
$\mathbf{k}_{\text{M}}$. The efficiency $\eta$ gives the ratio of the photon flux in L leaving volume ${\mathcal{V}}$ over
the photon flux in M entering ${\mathcal{V}}$. As shown in Fig.~\ref{fig4}(d), the conversion efficiency is approximately
$\eta \approx 0.3\%$ over a range of low intensities and then decreases with increasing $I_\text{M}$.
Note that $\eta$ in Eq.~(\ref{eta}) is a measure of the efficiency of the physical conversion
process in the Rydberg medium based on the
microwave power $I_{\text{M}} S_{\text{M}}$  impinging on $S_{\text{M}}$. This power is smaller than the total power emitted by the horn antenna since the M field has not been focused on $\mathcal{V}$ in our setup.

The good agreement between our model and the experimental data allows us to theoretically explore other geometries.
To this end we consider that the microwave fields M and A are co-propagating with the P field,
and assume that all other parameters are the same~\cite{Note1}.
We numerically evaluate the generated light power $P_{\text{L}}^{\parallel}$ for this setup and calculate the efficiency  $\eta^{\parallel}$
by replacing $P_{\text{L}}$ with $P_{\text{L}}^{\parallel}$ and  $S_{\text{M}}$ with $S_{\text{M}}^{\parallel}=\pi w_{\text{P}}^2$ in Eq.~(\ref{eta}).
We find $\eta^{\parallel} \approx 26\%$, which is approximately two orders of magnitude larger than $\eta$. This increase is mostly due to the
geometrical factor $S_{\text{M}}/S_{\text{M}}^{\parallel}\approx 91$, since $P_{\text{L}}^{\parallel} \sim P_{\text{L}}$. Note that such a value
for $\eta^{\parallel}$ is consistent with the efficiency achieved by a similar near-resonance frequency
mixing scheme in the optical domain~\cite{merriam2000}.
%

In conclusion, we have demonstrated coherent microwave-to-optical conversion via
a six-wave mixing process utilizing the strong coupling of electromagnetic
fields to  Rydberg atoms. We have established the coherence of the conversion by
a heterodyne measurement and  demonstrated a large bandwidth by measuring the generated light
as a function of the input microwave frequency.
Coherence and large bandwidth are essential for taking our scheme
to the single photon level and using it in quantum technology applications. Our
results are in  good agreement with theoretical simulations based on an independent
atom model thus showing a limited impact of atom-atom interaction on
our conversion scheme.

This work  has focussed on  the physical conversion mechanism in Rydberg systems and
provides several possibilities for future studies and applications.
Alkali atom transitions offer a wide range of frequencies in the optical and microwave
domain with properties similar to those exploited in this work. For example, the
conversion of a microwave field to  telecommunication wavelengths is
possible by switching to different optical transitions and/or using different atomic
species~\cite{gilbert1993frequency,bouchiat1989cs,jacobs:17}, which makes our approach
promising for classical and quantum communication applications.
Moreover, it has been theoretically shown that bidirectional conversion with near-unit efficiency is possible by using a different Rydberg excitation scheme and well-chosen detunings of the auxiliary fields~\cite{kiffner:16}. Such non-linear conversion with near-unit efficiency has only been experimentally realized in the optical domain~\cite{merriam1999}. Reaching this level of efficiency requires good mode-matching between the millimeter waves and the auxiliary optical fields~\cite{kiffner:16}, which can be achieved either by tightly focusing the millimeter wave, or by confining it to a waveguide directly coupled to the conversion medium~\cite{Hafezi:2012, Hogan2012}. Eventually, extending our conversion scheme to millimeter waves in a cryogenic environment~\cite{Hermann2014,cano:11} would pave the way towards quantum applications.\\
%

\begin{acknowledgments}
The authors thank Tom Gallagher for useful discussions and acknowledge the
support by the National Research Foundation, Prime
Ministers Office, Singapore and the Ministry of Education, Singapore under the
Research Centres of Excellence programme. This work is supported by Singapore
Ministry of Education Academic Research Fund Tier 2 (Grant No.
MOE2015-T2-1-085). M.K. would like to acknowledge the use of the University of
Oxford Advanced Research Computing (ARC) facility
(http://dx.doi.org/10.5281/zenodo.22558).
\end{acknowledgments}

%
%
%

%

%
%
%
\end{document}